\documentclass[twocolumn,pre,aps,showpacs,amsmath,amssymb,superscriptaddress]{revtex4-1}

\usepackage{hyperref}
\usepackage{amsmath,amssymb}
\usepackage{amsfonts,amsthm}
\usepackage{graphics}
\usepackage{graphicx}
\usepackage{dcolumn}
\usepackage{color}
\usepackage{bm}

\usepackage[normalem]{ulem}

\begin{document}

%\title{Anticipated synchronization due to GABAergic inhibitory autapic connection}
%\title{ Inhibitory autapses in neurons induces faster internal dynamics}
%\title{ Inhibitory autapse increases the frequency of neuronal spikes: a dynamical approach via non-intuitive synchronization}
%\title{ Inhibitory autapse may induce non-intuitive behaviour of neurons and microcircuits}

\title{Neuronal heterogeneity modulates phase-synchronization between unidirectionally coupled populations with excitation-inhibition balance}

\author{Katiele V. P. Brito}
\author{Fernanda S. Matias}
\thanks{fernanda@fis.ufal.br}
\affiliation{Instituto de F\'{\i}sica, Universidade Federal de Alagoas, Macei\'{o}, Alagoas 57072-970 Brazil}

\begin{abstract}

Several experiments and models have highlighted the importance of neuronal heterogeneity in brain dynamics and function. However, how such a cell-to-cell diversity can affect cortical computation, synchronization, and neuronal communication is still under debate. Previous studies have focused on the effect of neuronal heterogeneity in one neuronal population. 
Here we are specifically interested in the effect of neuronal variability on the phase relations between two populations, which can be related to different cortical communication hypotheses.
It has been recently shown that two spiking neuron populations unidirectionally connected in a sender-receiver configuration can exhibit anticipated synchronization (AS), which is characterized by a negative phase-lag. This phenomenon has been reported in electrophysiological data of non-human primates and human EEG during a visual discrimination cognitive task. 
 In experiments, the unidirectional coupling could be accessed by Granger causality and
 can be accompanied by both positive or negative phase difference between cortical areas. 
 Here we propose a model of two coupled populations in which the neuronal heterogeneity can determine the dynamical relation between the sender and the receiver and can reproduce phase relations reported in experiments. 
 Depending on the 
distribution of
parameters characterizing the neuronal firing patterns, the system can
exhibit both AS and the usual delayed synchronization regime (DS, with positive phase)
as well as a zero-lag synchronization regime and phase bistability between AS and DS.
 Furthermore, we show that our network can present diversity in their phase relations maintaining the excitation-inhibition balance.

% In neuronal models a transition from the usual delayed synchronization (DS, with positive phase-lag) 
%to AS was shown to occur depending on the relationship between excitatory and inhibitory synaptic conductances. that a local property of the receiver population as neuronal heterogeneity can determine the dynamical relation between the sender and the receiver.  Depending on the  distribution of parameters characterizing the neuronal firing patterns, the system can exhibit both AS and DS as well as a zero-lag synchronization regime and phase bistability between AS and DS. Furthermore, we show that our network can present diversity in their phase relations maintaining the excitation-inhibition balance.

\end{abstract}
%\pacs{87.18.Sn, 87.19.ll, 87.19.lm}
\maketitle

\section{Introduction}

The coherent activity of different cortical areas has been considered related to plenty of cognitive functions, such as, object recognition, visual-motor integration, and working memory ~\cite{Varela01,Buzsaki06}. Many hypotheses about neuronal communication take into account the role of brain oscillation and phase synchronization, for example:
the binding by synchrony~\cite{Singer99}
the communication through coherence~\cite{Fries05,Bastos15}
the gating by inhibition~\cite{Jensen10}, and 
nested oscillations~\cite{Bonnefond17}.
Although the mechanisms involved in large-scale integration are still unknown, they have been extensively studied with biologically inspired neuronal population models~\cite{Wang10}.

Spiking neurons are typically considered the building blocks of the population dynamics and neuronal diversity is ubiquitous across the nervous system. However, the functional significance of neuronal heterogeneity is still under investigation~\cite{Marsat10,Padmanabhan10,Savard11,Angelo12,Tripathy13}. 
From the experimental point of view, empirical observations show considerable variability in the response properties of different neurons~\cite{Bealieu93,Markram04,Angelo12}, and its relation with efficient neural coding~\cite{Marsat10}.
From the modeling side, a variety of computational network models of one neuronal population have been employed to study the effect of heterogeneity in synchronization and coding capabilities~\cite{Golomb93,Shamir06,Perez10,Mejias12,Mejias14,Rossi2020}, in self-sustained activity~\cite{Borges2020}, and in persistent activity, which could underlie the cognitive function of working memory~\cite{Renart03,Hass19}.

The previous studies mentioned above have investigated the effect of neuronal heterogeneity in one neuronal population. 
Here we are interested in the role of heterogeneity in the phase synchronization of a physiologically plausible model of two neuronal populations unidirectionally coupled. This could be potentially useful to understand the influence of neuronal variability on the communication between distant brain areas~\cite{Singer99,Fries05,Bastos15,Jensen10,Bonnefond17}.
The neuronal population model studied here have been employed before, in the light of anticipated synchronization ideas~\cite{Voss00,Matias14}, to explain electrophysiological results in non-human primate showing that unidirectional Granger causality can be accompanied by both positive or negative phase difference between cortical areas~\cite{Matias14,Montani15,Brovelli04,Salazar12}. However, the effect of neuronal variability on those phase relations have not been explored.

The typical synchronized regime between two unidirectionally coupled systems exhibits a positive phase-lag in which the sender is also the leader. This regime is usually called delayed synchronization (DS) or simply lagged synchronization.
However, it has been shown that the sender-receiver configuration can also synchronize with a negative phase-lag if the system can be described by the following equations in which the receiver is subjected to a delayed self-feedback~\cite{Voss00}:
\begin{eqnarray}
\label{eq:voss}
\dot{\bf {S}} & = & {\bf f}({\bf S}(t)), \\
\dot{\bf {R}} & = & {\bf f}({\bf R}(t)) + {\bf K}[{\bf S}(t)-{\bf R}(t-t_d)]. \nonumber 
\end{eqnarray}
The stable solution ${\bf R}(t)={\bf S}(t-t_d)$, 
characterizes the counter-intuitive situation in which the receiver leads the sender and it is called anticipated synchronization (AS). This means that the activity of the receiver predicts the activity of the sender by an amount of time $t_d$.
In the last 20 years, this solution have been extensively studied in physical systems both theoretically~\cite{Voss00,Voss01b,Voss01a,Ciszak03,Masoller01,HernandezGarcia02,Sausedo14} and
experimentally~\cite{Sivaprakasam01,Ciszak09,Tang03}. In particular, AS in unidirectional circuits have been employed for control and prediction of undesirable events~\cite{Ciszak09}.

It has been shown that AS can also occur if the delayed self-feedback is replaced by different parameter mismatches at the receiver~\cite{Kostur05,Pyragiene13,Simonov14}, a faster internal dynamics of the receiver~\cite{Hayashi16,Dima18,Pinto19,DallaPorta19}, as well as by an
inhibitory loop mediated by chemical synapses~\cite{Matias11,Matias16,Matias17,Mirasso17}.
 Moreover, when the feedback is not hard-wired in the equation but emerges from system dynamics, two neuronal population can present phase diversity and a transition from AS to DS through zero-lag synchronization, induced by synaptic properties~\cite{Matias14,DallaPorta19}. Different patterns of phase synchronization can also emerge due to the time-delays in heterogeneous networks~\cite{Petkoski18}.
 More recently, it has been shown that a sender-receiver network can also present phase bistability between DS and AS regimes~\cite{Machado2020}. 
 It is worth to mentioning that in symmetric bidirectional coupled systems it is not possible to separately classify AS or DS, but only a lagged synchronization. 
 In case of asymmetric bidirectional influence we can define the stronger direction of influence as the sender-receiver direction. For example, the effect of a bidirectional connection in the AS regime have been studied in neuronal motifs coupled by chemical synapses~\cite{Matias16}. Since chemical synapses are unidirectional by their own nature, in such circuits each direction of influence is independent of each other.

Here we show that the neuronal heterogeneity can promote diversity of phase relations between two neuronal networks with excitation-inhibition balance. As far as we know, this is the first verification that the neuronal spiking properties of the population can promote AS and phase-bistability. 
In Sec.~\ref{model}, we describe the neuronal population model as well as the parameters that we use to change neuronal heterogeneity.
In Sec.~\ref{results}, we report our results, showing that the motif can exhibit phase-locking regimes: with positive, negative, and zero phases and a bistable regime which alternates from AS to DS. We also show that the excitatory and inhibitory conductances remain with a fixed relationship during oscillations.
Concluding remarks and a brief discussion of the significance of our findings for neuroscience are presented in Sec.~\ref{conclusions}.

\section{Network model with neuronal heterogeneity}
%Model of two unidirectionally coupled neuronal networks}
\label{model}

\begin{figure}[!ht]%
\includegraphics[width=0.7\columnwidth,clip]{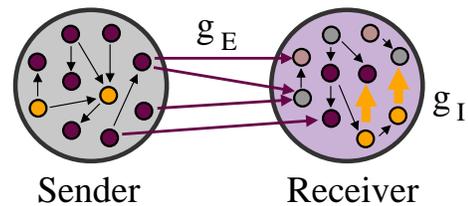}
\caption{\label{fig:motif}   (a) Schematic representation of two cortical areas, composed of hundreds of spiking neurons and chemical synapses, coupled in a sender-receiver configuration. Both populations have neuronal heterogeneity but for the Receiver (R) population we vary the distribution of
parameters determining the neuronal firing patterns.
The inhibitory feedback is controlled by the synaptic conductance $g_I$ at R, whereas the sender-receiver coupling is determined by excitatory synaptic conductances with $g_E$.
}
\end{figure}%

\subsection*{Each node is a neuron model with a specific firing pattern}

Our neuronal motif is composed of two unidirectionally coupled cortical-like 
neuronal populations: a sender (S) and a receiver (R),
see Fig.~\ref{fig:motif}(b). Each one is composed of 400 excitatory and 100 inhibitory
neurons~\cite{Matias14} which is the typical employed proportion of $80\%$ excitatory neurons and $20\%$ inhibitory neurons, based on anatomical estimates for neocortex~\cite{Brunel00}. 
Each neuron is
described by the Izhikevich
model \cite{Izhikevich03}:
\begin{eqnarray}
  \frac{dv}{dt} &=& 0.04v^2+5v+140-u +\sum_x I_{x}, \label{dv/dt}\\
  \frac{du}{dt} &=& a(bv-u). \label{du/dt}
\end{eqnarray}
In Eqs.~\ref{dv/dt} and~\ref{du/dt} $v$ is the membrane potential and
$u$ the recovery variable which accounts for activation
of K$^+$ and inactivation of Na$^+$ ionic currents.  $I_x$ are the synaptic currents provided by
the interaction with other neurons and external inputs.  If
$v\geq30$~mV, $v$ is reset to $c$ and $u$ to $u+d$.  

\begin{figure}[!t]%
\includegraphics[width=0.9\columnwidth,clip]{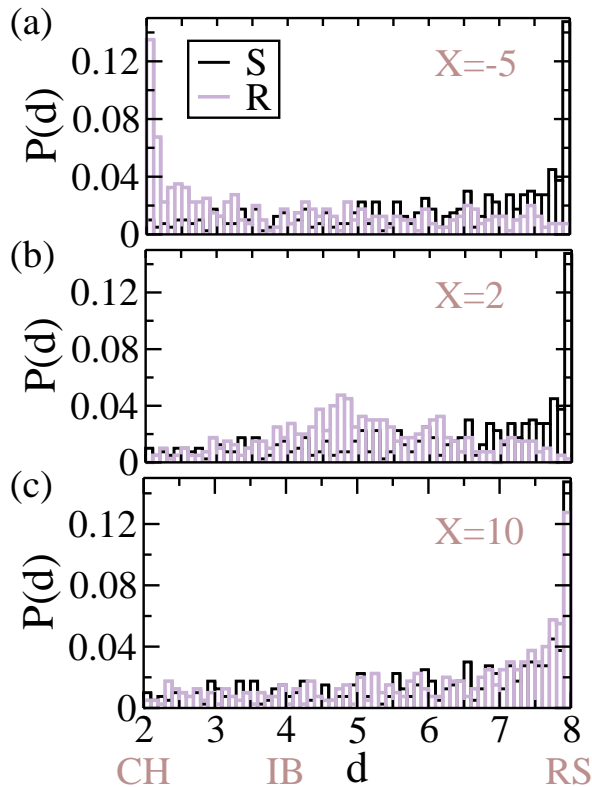}
\caption{\label{fig:variability}
Probability distribution of finding an excitatory neuron at R population with model parameter $d$, which determines its 
spiking firing pattern (see Eqs.~\ref{eq:c} and~\ref{eq:d}). (a) For $X=-5$ it is very probable to find $d=2$ (and consequently $c=-50$) which ensures that there are more chattering (CH) neuron in the population than other neuron types. (b) For $X=2$ the distribution of $d$ has a peak around $d=5$ and there are more intrinsically bursting neurons (IB, $d=4$ and $c=-55$) than other kinds. (c) For $X=10$ the majority of neurons are regular spiking (RS, $d=8$ and $c=-65$).}
\end{figure}

\begin{figure}[!ht]%
\includegraphics[width=0.96\columnwidth,clip]{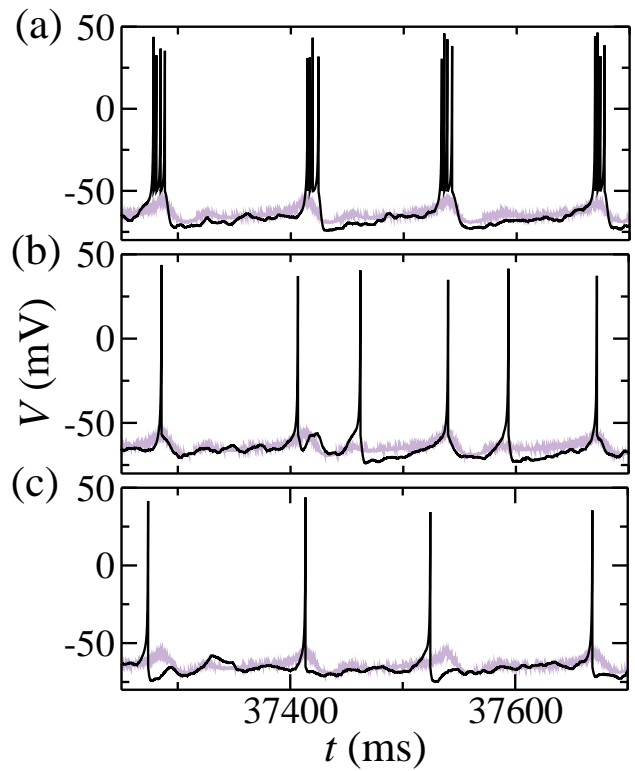}
\caption{\label{fig:neurons} Three exemplar neuronal time series to illustrate neuronal heterogeneity embedded in the network. Firing pattern during the oscillatory activity of (a) a chattering neuron (CH, $c=-50.4$), (b) an intrinsically bursting neuron (IB, $c=-56.9$), and (c) a regular spiking neuron (RS, $c=-64.9$). The purple line shows the mean activity of all neurons in the same population (for fixed $X=2$ and $g_I=2.0$~nS).} 
% $c=-50.352099 -56.901879 -64.882641$}
\end{figure}

To study the effects of neuronal variability at the receiver population, we use different values of (c,d) for the excitatory neurons in R:
\begin{eqnarray}
  c &=& -55-X+[(5+X)\sigma^2]-[(10-X)\sigma^2], \label{eq:c} \\
  d &=& 4  +Y-[(2+Y)\sigma^2]+[(4-Y)\sigma^2]. \label{eq:d}
\end{eqnarray}
We use $Y=2X/5$ to guarantee that neuronal type distribution is along the line changing from regular spikes (RS, $c=-65$ and $d=8$) to chattering (CH, $c=-50$ and $d=2$) via intrinsically bursting (IB, $c=-55$ and $d=4$). As we vary $X$ from $-5$ to $10$ the amount  of neuron of each type in the R population changes as shown in Fig.~\ref{fig:variability}. We vary $X$ throughout the paper to show how heterogeneity can affect phase relations between the two populations. Illustrative examples of neuronal firings embedding in the neuronal population but with different parameters are shown in Fig.~\ref{fig:neurons}.
Equations were integrated with the Euler method and a time step of $0.05$~ms.

\subsection*{Each link is an excitatory or inhibitory chemical synapse}

The connections between neurons in each population are assumed to be
fast unidirectional excitatory and inhibitory chemical synapses
mediated by AMPA and GABA$_\text{A}$. The synaptic currents are given
by:
\begin{equation}
\label{Ix}
I_{x} = g_{x}r_{x}(v-V_x),
\end{equation}
where $x=E,I$ (excitatory and inhibitory mediated by AMPA and
GABA$_\text{A}$, respectively), $V_E=0$~mV, $V_I=-65$~mV, $g_{x}$ is the 
synaptic conductance maximal strengths  and $r_{x}$
is the fraction of bound synaptic receptors whose dynamics is given
by:
\begin{equation}
\label{drdt}
  \tau_x\frac{dr_{x}}{dt}=-r_{x} + D \sum_k \delta(t-t_k),\\
\end{equation}
where the summation over $k$ stands for pre-synaptic spikes at times
$t_k$. D is
taken, without loss of generality, equal to $0.05$. The time decays are $\tau_{E}=5.26$~ms $\tau_{I}=5.6$~ms.  Each
neuron is subject to an independent noisy spike train described by a
Poisson distribution with rate $R$. The input mimics excitatory synapses 
(with conductances $g_{P}$) from $n$ pre-synaptic neurons external to the population,
each one spiking with a Poisson rate $R/n$ which, together with a constant external current $I_c$,  
determine the main frequency of mean membrane potential of each population. Unless otherwise stated, we have employed $R=2400$~Hz and $I_c=0$.

\begin{figure}[!h]%
\includegraphics[width=0.96\columnwidth,clip]{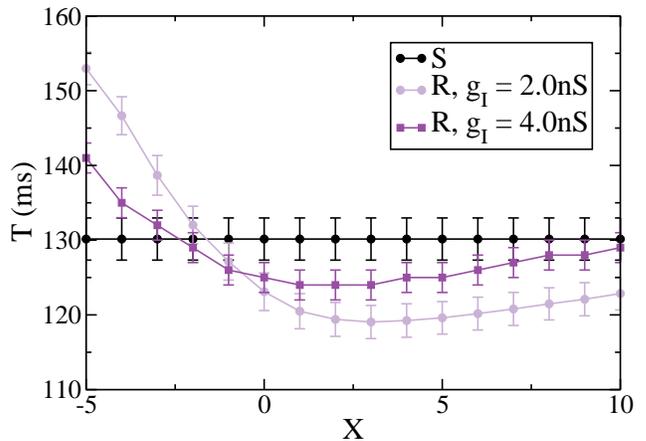}
\caption{\label{fig:onepopulation} Effect of neuronal heterogeneity in the oscillatory period of one population. The S and R populations are uncoupled ($g_E=0$~nS), and by changing $X$, we only change the neuronal parameters $c$ and $d$ of neurons in the R population (see Eqs.~\ref{eq:c} and~\ref{eq:d}). The oscillatory period of R depends on both inhibition and the neuronal variability parameter $X$. }
\end{figure}%

\begin{figure*}[!ht]%
\includegraphics[width=0.96\textwidth,clip]{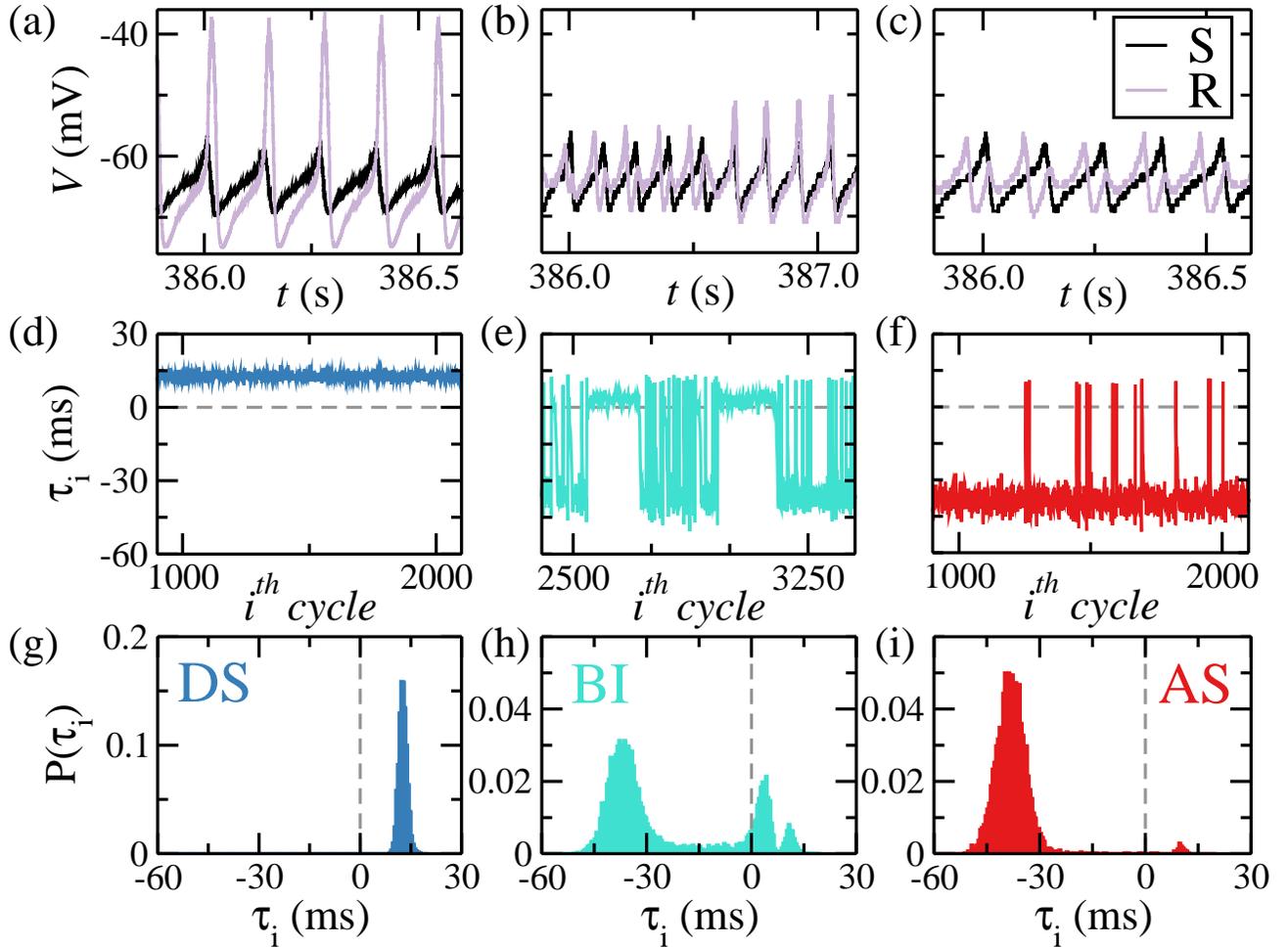}
\caption{\label{fig:DSBIAS} Characterizing the three different dynamical regimes exhibited by the two neuronal populations depending on the parameter $X$ controlling the neuronal spiking variability. There are two possible phase-locking regimes: delayed synchronization (DS, $X=-5$) and anticipated synchronization (AS, $X=10$), as well as a bistable regime in respect to the phase differences (BI, $X=2$).
(a)-(c) Time series for sender and receiver populations.
(d)-(f) Time delay $\tau_i$ per cycle.
(g)-(i) Probability distribution of time delays. (g) The unimodal distribution with a positive average characterizes DS, whereas (i) the negative average represents AS. (j) The bimodal distribution together with an alternation between cycles of DS randomly followed by cycles of AS characterizes the phase bistability.
Synaptic conductances are kept fixed: $g_E=0.5$~nS and $g_I=2.0$~nS.}
\end{figure*}%

Connectivity within each
population randomly targets 10\% of the neurons, with excitatory
conductances set at $g^{S}_{E} =g^{R}_{E} = 0.5$~nS. Inhibitory conductances are fixed at the sender population $g^{S}_{I} = 4.0$~nS 
and $g_I$ at the receiver population is varied throughout the study (see Fig.~\ref{fig:motif}).
Each neuron at the R population receives 
20 fast synapses (with conductance $g_{E}$) from random excitatory
neurons of the S population.
It is worth emphasizing that to analyze the 
excitation-inhibition balance at each neuron, we use $G_x=g_{x}r_{x}$ as the effective synaptic conductance at each time step, which is a fraction of the maximal possible value of $g_{x}$.

\subsection*{Characterizing phase relations between sender and receiver populations}
Since the mean membrane potential $V_x$ ($x=$S, R) of each population 
(which we assume as a crude approximation of the measured LFP) is
noisy, we average within a sliding window of width 5-8~ms to obtain a smoothed signal, from which we can extract 
the peak times $\{t^{x}_{i}\}$ (where $i$ indexes the peak). The
period of a given population in each cycle is thus $T^{x}_i \equiv t^{x}_{i+1}-t^{x}_{i}$. 
For a sufficiently long time series, we compute the mean period
$T_x$ and its variance.  

In a similar way we calculate the time delay in each cycle
$\tau_i=t^{R}_{i}-t^{S}_{i}$. Then, if $\tau_i$ obeys a unimodal distribution,  we calculate $\tau$ as the mean
value of $\tau_i$ and $\sigma_{\tau}$ as its variance.  
If $T_S  \approx  T_R $
and $g_E$ is independent of the initial conditions, the populations exhibit
oscillatory synchronization with a phase-locking regime. We indistinguishably use the term phase difference or time delay since it is always possible to associate both: $\phi_i=2\pi \tau_i/ T_{S}$. In all those calculations we discard the
transient time.  

It is also possible to estimate the time-delay by using the peak of 
the delayed cross-correlation function $C(V_S,V_R,\Delta t)$ between the the mean membrane potential of the S and R
populations. This function can be calculated as: 
\begin{equation}
C(V_S,V_R,\Delta t)= \frac {(\sum{V_{S}^{i}- \overline{V_S}})( \sum{V_{R}^{i+\Delta t}- \overline{V_R}})} {\sqrt{\sum{(V_{S}^{i}- \overline{V_S})^2}}  \sqrt{\sum{(V_{R}^{i}- \overline{V_R})^2}}} \; .
\label{eq:crosscorr}
\end{equation} 
The values of $C(V_S,V_R,\Delta t)$ also indicate the level of synchronization between the two populations.

\section{Results}
\label{results}

\subsection*{The effect of neuronal heterogeneity in one population}

To study the effect of neuronal heterogeneity in the oscillatory properties of only one population, 
we analyze the uncoupled case ($g_E=0$). For comparison, the Sender population, which has a fixed distribution of neuronal spiking types, oscillates with a mean period of $T_S=130$~ms. As we change the neuronal types distribution of the Receiver population by varying $X$ in Eq.~\ref{eq:c}, the oscillatory period of R varies from more than $150$~ms to less than $120$~ms (see Fig.~\ref{fig:onepopulation} for two different values of inhibition: $g_I=2.0$~nS and $g_I=4.0$~nS ). A decrease in the number of chattering neurons and an increase in the number of intrinsically bursting neurons facilitate faster oscillations.

This means that we can control the internal dynamics of the population by changing the local properties of the neurons. In particular, we can turn the receiver faster maintaining the excitation-inhibition balance by changing $X$.
It is worth mentioning that the E/I balance has been extensively related to network dynamics, information processing in the nervous system, and social dysfunction~\cite{Brunel03,Landau16,Yizhar11}.
Moreover, it was not always possible to keep the E/I balance in previous studies in which the internal dynamics of the receiver were determined by the relatioship between inhibitory and excitatory synaptic currents at the receiver~\cite{DallaPorta19,Machado2020}.

\subsection*{Local properties at the receiver population modulates global phase relations between two population}

The free-running properties of each population can influence the synchronization patterns between them, when we turn the sender-receiver coupling on ($g_E>0$)~\cite{DallaPorta19,Pinto19,Dima18}. In fact, for $g_E=0.5$~nS and $g_I=2.0$~nS the motif can exhibit three different regimes depending on $X$. Illustrative examples of these dynamics are shown in Fig.~\ref{fig:DSBIAS} for $X=-5$, $X=2$ and $X=10$. The distribution of neuronal spiking parameters shown in Fig.~\ref{fig:variability} 
ensures that the receiver population exhibits more chattering neurons for $X=-5$, more intrinsically bursting neurons for $X=2$, and more regular spiking neuron for $X=10$.
For $X=-5$ the system presents the usual phase-locking regime with positive mean time delay $\tau=13$~ms. This means that a peak of the mean membrane potential of the sender population is followed by a peak of the receiver.
Therefore, the time delay in each cycle $\tau_i$ is positive and fluctuates around a well defined mean value (see the left column in Fig.~\ref{fig:DSBIAS}). For this example, the delayed cross-correlation function $C(V_S,V_R,\Delta t)$ calculated by Eq.~\ref{eq:crosscorr} has a local peak of $0.92$ for $\Delta t=15$~ms.

For $X=10$, a peak of the Sender is, for the majority of the cycles, preceded by a peak of the receiver ($\tau_i<0$) and, consequently, the mean time delay is negative  (see the right column in Fig.~\ref{fig:DSBIAS}). This characterizes the anticipated synchronization regime (AS, with mean time delay $\tau=-39$~ms for this example).
 This counter intuitive regime explains the observed unidirectional influence with negative phase difference verified in LFP monkey data~\cite{Brovelli04,Salazar12,Matias14} as well as in human EEG~\cite{Carlos2020}. Moreover, AS could be possibly related to commonly reported short latency in visual systems~\cite{Orban85,Nowak95,Kerzel03,Jancke04,Puccini07,Martinez14}, olfactory circuits~\cite{Rospars14}, 
songbirds brain~\cite{Dima18} and human perception~\cite{Stepp10,Stepp17}. For this example, the delayed cross-correlation function $C(V_S,V_R,\Delta t)$  has a local peak of $0.84$ for $\Delta t=-39$~ms.

\begin{figure}[t]%
\includegraphics[width=0.99\linewidth,clip]{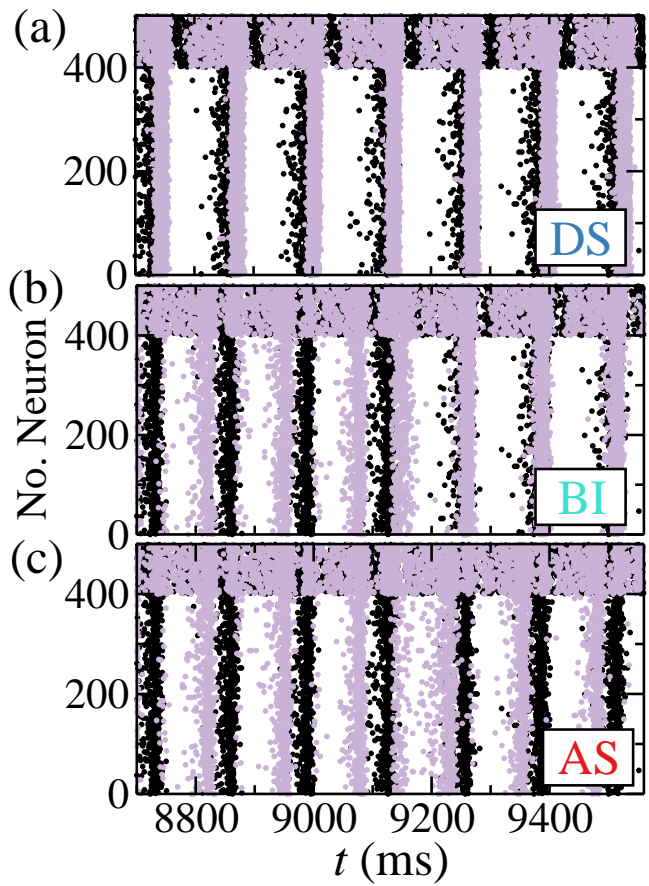}
\caption{\label{fig:raster} Illustrative examples of the raster plots for all neurons in each dynamical regime. Black dots are neurons from the sender population and purple (light gray) dots are neurons at the receiver. All parameters as in Fig.~\ref{fig:DSBIAS} and we only change the distribution of CH, IB, and RS spiking neurons in the network: $X=-5$ generates DS, $X=2$ promotes the phase bistability, and for $X=10$ the motif exhibits AS.
}
\end{figure}%

For intermediate values of $X$ the system exhibits a phase-bistability between these two possible phase-locking regimes: DS and AS (see Fig.~\ref{fig:DSBIAS}(b), (e) and (h) with $X=2$). The activity alternates from a few periods of DS eventually followed by a few periods of AS.
In other words, the time delay $\tau_i$ is positive for a few cycles, with a well-defined mean value and standard deviation (close to $\tau=4.5$~ms in this case), which is similar to a DS regime for a certain amount of periods. 
Then, the system randomly switches to a different dynamics in which $\tau_i$ is negative during a few other cycles (the mean time delay is close to $\tau=-36$~ms for this example).
Therefore, in this regime, the system cannot be simply characterized by the mean time delay ($\tau$).

A phase-bistability between AS and DS has been previously reported for very small values of inhibition~\cite{Machado2020}. It has been proposed that this phase bistability could be a plausible model for differences in phase synchronization of MEG data during bistable perception~\cite{Kosem14}.
Kosem et al.~\cite{Kosem14} have shown that when participants listening to bistable (or ambiguous) speech sequences that could
be perceived as two distinct word sequences repeated over
time, their MEG recordings
present phase-differences related to which sequence they are perceiving. Therefore, we propose that this phase-bistability can also be promoted by neuronal variability in cortical regions.

\begin{figure}[!t]%
\includegraphics[width=0.99\columnwidth,clip]{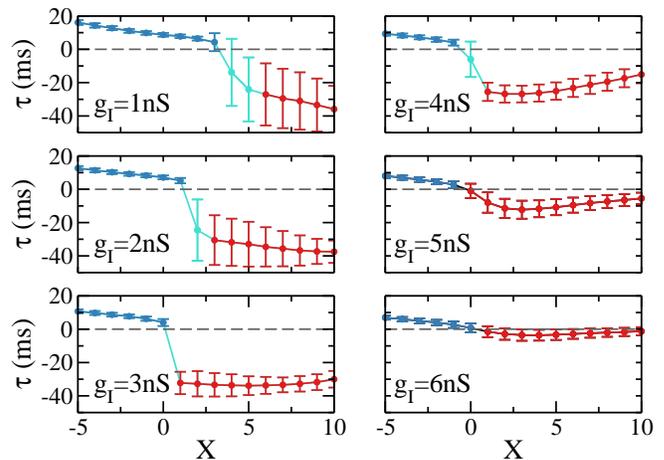}
\caption{\label{fig:tauvsX} Mean time delay $\tau$ between neuronal population as a function of the parameter controlling the neuronal heterogeneity $X$. The DS-AS transition can occur via a bistable regime o through zero-lag synchronization for sufficient large inhibition ($g_I=5$~nS and $g_I=6$~nS).}
\end{figure}%

The three different regimes can also be visualized in a raster plot as in Fig.~\ref{fig:raster}. In these plots, each dot represents a spike of the i-th neuron indexed by the vertical axis at the time indicated by the horizontal axis. It is worth mentioning a few points. Interneurons (indices 400 to 499) are spiking with higher frequency than excitatory neurons (indices 0 to 399) in both populations.
The spiking pattern of the excitatory neurons at the receiver population in the DS regime is more concentrated close to the peak of the mean activity. This means that the neurons are more synchronized with each other. 
On the other hand, the variability along the entire period is larger in the AS regime. 
Bursting neurons and regular spiking neurons can more easily fire in the middle of the period than chattering neurons (see an example in Fig.~\ref{fig:neurons}). 
Since we are changing the proportion of CH, IB, and RS neurons to obtain the transition from one regime to another, we could speculate that CH neurons facilitate the synchronization of the entire population and the usual DS regime, whereas RS neurons allow the observed larger diversity and the AS regime.

\begin{figure}[!t]%
\includegraphics[width=0.99\columnwidth,clip]{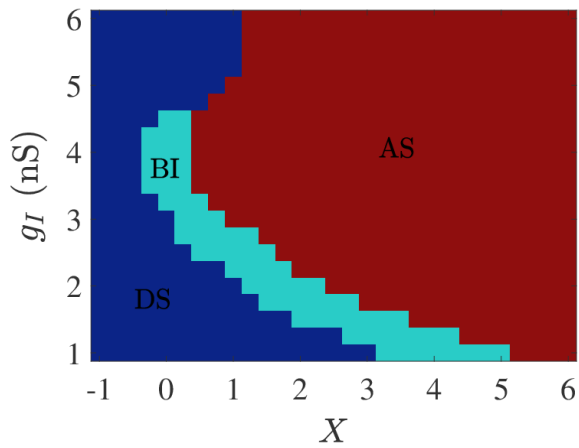}
\caption{\label{fig:3Dplot}  Scanning parameter space to visualize the phase relation as a function of inhibitory conductance $g_I$ and the parameter controlling neuronal variability $X$. The three regimes: DS, AS and BI can be verified for a reasonable set of parameters. Horizontal lines for integer vales of $g_I$ are shown in Fig.~\ref{fig:tauvsX}.
}
\end{figure}

By incremental changes in the neuronal heterogeneity, the system can undergo a transition from DS to AS through the bistable regime or via zero lag-synchronization depending on the amount of inhibition at the receiver.
Fig.~\ref{fig:tauvsX} shows the mean time delay as a function of $X$, the parameter controlling the neuron type distribution, for different values of inhibitory conductance $g_I$. Each curve corresponds to a horizontal line in Fig.~\ref{fig:3Dplot} which displays a two-dimensional projection of the parameter space of our model ($X$,$g_I$). 

For $g_I>4.5$nS the transition from DS to AS is continuous and we can find virtually any value of mean time delay between the two populations, including a zero-lag synchronization, in which the peak of activity of both regions occurs very close on average (see $g_I=5$nS and $g_I=6$nS in Fig.~\ref{fig:tauvsX}). In Fig.~\ref{fig:3Dplot} we can see that the end of the bistable regime between DS and AS, gives place to a DS-AS transition via zero-lag, and coincides with a change in the slope of the boundary between DS and AS. This re-entrant behavior allows the system to have, for example for fixed $X=1$, a first DS-AS transition via bistability followed by an AS-DS transition via zero-lag as a function of $g_I$ (this would corresponds to a vertical line $X=1$ in Fig.~\ref{fig:tauvsX}).

The zero-lag regime has been extensively studied as a non-intuitive regime that can overcome the synaptic delays between distant areas~\cite{Vicente08,Gollo14}. The first experimental results about the total synchronization of distant neurons originated many theories about neuronal communication as binding by synchrony~\cite{Singer99} and communication through coherence~\cite{Fries05}. With new experimental results about phase relation diversity, the latter hypothesis has been adapted to include the effect of non-zero phase relations~\cite{Bastos15,Maris13,Maris16}. 
Furthermore,  it has been shown that 
in unidirectionally coupled cortical populations
the DS-AS transition via zero-lag can be mediated by synaptic conductances~\cite{Matias14,DallaPorta19} and that
spiking-timing dependent plasticity can promote auto-organized zero-lag synchronization ~\cite{Matias15}. 

It is worth to emphasizing that previous experimental studies with brain signals have shown that unidirectional influence can be accompanied by positive, zero or negative phase differences in different frequency bands ~\cite{Brovelli04,Salazar12,Matias14,Carlos2020}.
In our model, the populations oscillate with a period of $130$~ms which is equivalent to a frequency close to $7.7$~Hz. This means that the time delay of $13$~ms in the DS regime shown in Fig.~\ref{fig:DSBIAS}(a) is equivalent to a phase delay of $0.2\pi$~rad, whereas the anticipation in Fig.~\ref{fig:DSBIAS}(c)  has a phase delay of $-0.6\pi$~rad. 
Therefore our model is able to reproduce unidirectional influence in the alpha band with both positive, zero and negative phase relations. 
It has been recently shown that human EEG can present unidirectional influence between signals with both positive and negative phase relation~\cite{Carlos2020]}. All the 11 analyzed volunteers presented pairs of electrodes synchronized in the alpha band (from $7$ to $13$~Hz) with unidirectional influence and a diversity of phase relations including positive, negative and zero phase lags.
Two other studies with monkey LFP have shown that unidirectional influence can be accopanied by the couterintuitive negative phase difference reported here. In Brovelli et al.~\cite{Brovelli04} they have shown that somatosensory motor cortex can influence motor cortex during oscillatory activity with main frequency around $24$~Hz and a negative time difference of $-8.7$~ms.
In a different experiment~\cite{Salazar12}, it has been shown the posterior parietal cortex can unidirectinal influence the prefrontal cortex in a frequency of $17$~Hz and present a time difference that can range from $-2.45$~ms to $-6.53$~ms.

\subsection*{Excitation-Inhibition balance}
\label{balance}

In previous studies \cite{Matias14,DallaPorta19} with sender-receiver populations, the different regimes were achieved by changing the excitation-inhibition relation via modifications in the internal inhibition, the sender-receiver coupling, or the amount of external noise simulated as excitatory synapses~.
In particular, the bistable regime reported in ~\cite{Machado2020} required very small values of inhibitory conductance. Here we can obtain different phase relations ensuring that the excitation-inhibition balance is maintained by only changing the neuronal heterogeneity. 

\begin{figure}[t]%
\includegraphics[width=0.99\columnwidth,clip]{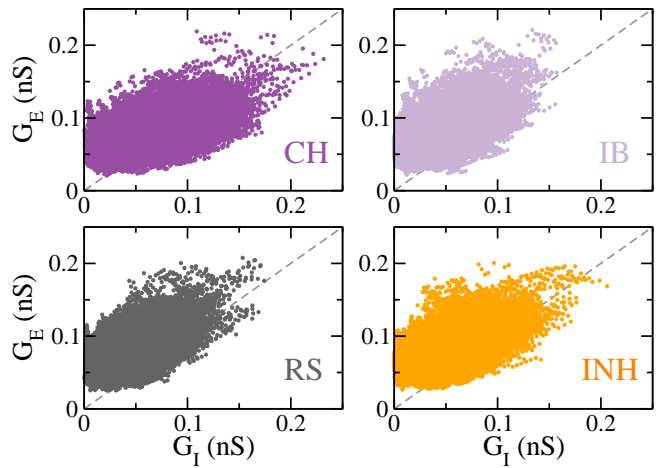}
\caption{\label{fig:EI}  
Excitatory-inhibitory balance for four illustrative neurons (three excitatory: CH, IB, RS and one inhibitory) at the receiver population during the bistable regime ($X=2$ and $g_I=2$~nS). Excitatory and inhibitory effective synaptic conductances preserve an approximate proportional relation to each other in the course of the oscillations. 
}
\end{figure}

We can see the excitatory-inhibitory balance in Fig.~\ref{fig:EI} where the two effective synaptic conductances are plotted
against each other for four different neurons of the receiver population: one chattering, one intrinsically bursting, one regular spiking, and an inhibitory neuron.
These plots yield a linear relationship, which means that the excitatory and inhibitory synaptic conductance increase and decrease in time in such a way that the ratio of the two $G_E/G_I$ remains almost fixed. This ratio slightly varies from neuron to neuron. The E/I balance has been related to a variety of dynamical features such as oscillatory activity~\cite{Brunel03,Compte03}, information processing and social dysfunction~\cite{Yizhar11} as well as cortical complexity~\cite{Barbero2020}.

\section{Concluding remarks}
\label{conclusions}

We have shown that neuronal heterogeneity can promote different phase relations between spiking neuron networks. We have verified that by changing the proportion of different types of neuronal firing patterns the system can present phase-locking regimes with positive (DS), negative (AS), and zero phase differences as well as a bistable regime. The DS-AS transition could possibly explain commonly reported short latency in visual systems~\cite{Orban85,Nowak95,Kerzel03,Jancke04,Puccini07,Martinez14}, olfactory circuits~\cite{Rospars14}, 
songbirds brain~\cite{Dima18} and human perception~\cite{Stepp10,Stepp17}, whereas the bistable regime can be associated with bistable perception during ambiguous stimulation ~\cite{Kosem14,Machado2020}. Previous studies on AS~\cite{Matias14,DallaPorta19} and phase-bistability~\cite{Machado2020} in the neuronal population have not explored the effects of neuronal properties on the network dynamics.
Moreover, the possibility of change the neuronal variability allows us to remain in the excitation inhibition balance which has been considered a fundamental property of the health brain~\cite{Brunel03,Yizhar11,Compte03,Barbero2020}.

Although examples of neuronal variability exist throughout the brain, their importance for the information process remains a source of debate. The functional roles of intrinsic neuronal heterogeneity on the network dynamics have been more extensively studied in the last years~\cite{Marsat10,Padmanabhan10,Savard11,Angelo12,Tripathy13}. Theoretically, the effect of heterogeneity on synchronization properties of only one neuronal network has been studied in different models~\cite{Golomb93,Perez10,Mejias12,Mejias14,Rossi2020}.
In particular, Rossi et al.~\cite{Rossi2020} argued that their results can be important in the light of communication through coherence ideas~\cite{Fries05,Bastos15}. Here we give a step further in this direction by showing the specific effect of heterogeneity in the phase synchronization patterns of two coupled networks. In such a case, we could investigate the effect of neuronal heterogeneity in phase relations and, consequently, in communication between distant cortical areas.

Differently from the first papers about AS~\cite{Voss00,Ciszak03}, here the anticipation time is not hard-wired in the dynamical equations, neither associated with the excitation-inhibition relation but rather emerges from the neuronal dynamics and heterogeneity.
This opens new possibilities to study how neuronal variability modulates the phase relation diversity between cortical areas which has been the object of more intense investigation in the last years~\cite{Maris13,Maris16}.
In particular, including effects from homeostasis and the variability of inhibitory neurons are natural next steps that we are
currently pursuing.

%\appendix
%\section{\label{Appendix}Appendix}

\begin{acknowledgments}
The authors thank FAPEAL, UFAL, CNPq (grant 432429/2016-6), and CAPES (grant 88881.120309/2016-01) for financial support.

\end{acknowledgments}
% 
% \pagebreak

\bibliography{matias}

\end{document}